\documentclass[aps
,12pt,tightenlines
,endfloats,twoside,showpacs]{revtex4}

\usepackage{graphicx}
\usepackage{epsfig}
\usepackage{amsmath}
\usepackage{amstext}
\usepackage{amssymb}
\usepackage[english]{babel}
\usepackage[latin2]{inputenc}
 \DeclareGraphicsExtensions{.pdf}
 \DeclareGraphicsExtensions{.eps,.ps}
\begin{document}
\author{R. Parzy\'nski, M. Sobczak and A. Pluci\'nska}
\affiliation{Faculty of Physics, Adam Mickiewicz University,
   Umultowska 85, 61-614 Pozna\'n, Poland}
\title{An iterative method for extreme optics of two--level systems}

\begin{abstract}
We formulate the problem of a two-level system in a linearly polarized
laser field in terms of a nonlinear Riccati-type differential equation
and solve the equation analytically in time intervals much shorter
than half the optical period. The analytical solutions for subsequent
intervals are then stuck together in an iterative procedure to cover
the scale time of the laser pulse. This approach is applicable to
pulses of arbitrary (nonrelativistic) strengths, shapes and durations,
thus covering the whole region of light-matter couplings from weak
through moderate to strong ones. The method allows quick insight into
different problems from the field of light--matter interaction. Very
good quality of the method is shown by recovering with it a number of
subtle effects met in earlier numerically calculated photon-emission
spectra from model molecular ions, double quantum wells, atoms and
semiconductors. The method presented is an efficient mathematical tool
to describe novel effects in the region of, e.g., extreme nonlinear
optics, i.e., when two--level systems are exposed to pulses of only a
few cycles in duration and strength ensuring the Rabi frequency to
approach and even exceed the laser light frequence.
\end{abstract}

\pacs{42.50.Md, 42.50.Hz, 42.65.Ky, 42.65.Re}

\maketitle

\section{Introduction}\label{int}

In the theory of light-matter interactions there is probably no more
        fundamental model than the two-level one \cite{allen}. Over
        the last decade, for example, the model has succedded in
        explaining the main features of propagation of strong a
        few-cycle pulses through atomic and semiconductor media
        \cite{ziolk, casp, hughes1, hughes2, kalosha, tara, weg1a,
        weg1b, xiao, cheng, weg2}, e.g., carrier wave Rabi flopping,
        third-harmonic generation in disguise of second harmonic and
        carrier-envelope phase effects, to name a few. It has also
        made a basis for the description of high-order harmonic
        generation from a single atom \cite{mil}, a symmetric
        molecular ion \cite{zuoa, zuob, ivanov1} and a double-well
        quantum structure \cite{ivanov2, bavli, levi} with emphasis on
        the strongly non-perturbative picture of the phenomenon and
        the occurrence of peaks in the spectrum of coherently
        scattered light at the positions of even harmonics. When
        applied to double wells, the model turned out to be successful
        also in explaining the effect of laser control of tunneling
        \cite{plata}.

Despite its dissemination in atomic, molecular and solid state physics
        the two-level model of light-matter interaction still suffers
        from the lack of exact analytical solution covering the whole
        range of laser intensities as well as pulse shapes and
        durations. The analytical solutions known hitherto cover only
        some different limiting cases. For instance, the most
        celebrated rotating-wave-approximation (RWA) solution
        \cite{allen} is restricted to laser intensities ensuring the
        resonance Rabi frequency, $\Omega_{R}$, to be much smaller
        than the laser frequency, $\omega $. Beyond RWA, the known
        analytical solutions include the non-RWA corrections along a
        perturbative procedure (e.g. \cite{genkin, parz1}), either are
        valid in the so-called multiphoton excitation region
        \cite{zuoa, zuob} ($\omega <<\omega_{21}$ along with
        $\Omega_{R}<< \omega_{21}$, where $\omega_{21}$ stands for the
        frequency separation between the two levels) or in the quite
        opposite strong coupling region \cite{zuoa, zuob, ivanov1,
        ivanov2, bavli, levi} ($\omega >> \omega_{21}$ and $\Omega_{R}
        >> \omega_{21}$). Probably, the only analytical solution
        covering the whole intensity region is the recent one of
        Tritschler, M\"ucke and Wegener \cite{weg3} for a box-shaped
        pulse, but obtained within the so-called square-wave
        approximation (SWA) consisting in replacing the actual time
        behavior of the field within half the optical cycle by a
        square of a constant appropriately chosen magnitude. Being
        approximate, this solution was able to reproduce only
        qualitatively some features of the exact numerical
        calculations, especially for the case of the resonant
        excitation ($\omega = \omega_{21}$), but was less convincing
        for distinctly off-resonant excitation.

The aim of our paper is to present a quick iterative procedure for the
        problem of the two-level system in linearly polarized laser
        field, based on an analytical solution of the Schr\"odinger
        equation in very short time intervals. The analytical solution
        turned out to be possible thanks to defining the problem of
        level populations in terms of a single nonlinear Riccati-type
        differential equation in conjunction with dividing each
        halfcycle of the pulse into a number of narrow slices of equal
        width and considering constant the electric field within each
        slice with a value determined by the pulse function at the
        middle of the slice. The analytically obtained solutions for
        all slices in the pulse are then stick together by a simple
        recurrence formula derived relating the boundary conditions in
        the adjacent slices. This approach offers a simple analytical
        formula for the ratio of level population amplitudes within
        each slice, resulting in equally simple analytical formulae
        for population inversion, induced dipole moment and spectrum
        of the radiation emitted by this dipole. The photon-emission
        spectra obtained along the above line reproduce the
        numerically calculated ones available in literature
        \cite{weg1a, weg1b, mil, zuoa, zuob, ivanov1, ivanov2, bavli,
        levi}. Moreover, our iterative method indicates weak points of
        the square-wave solution of Tritschler at el. \cite{weg3} and
        is proved to be particularly useful in the area of extreme
        nonlinear optics \cite{weg3}, i.e., when a few-cycle, strong
        pulses stimulate significant population dynamics in a
        two--level system on a time scale of half the optical cycle.

\section[The analitycal solution \ldots]{The analytical solution for short time intervals and iterative method}\label{anal sol}

When presenting our analytical solution for short time intervals we
start with the standard expansion $\psi (t) = b_{1}(t) |1 \rangle +
b_{2}(t) |2 \rangle $ for the wave function of the two-level system in
a laser field, where $|1 \rangle $ and $|2 \rangle $ stands for the
time-independent opposite-parity eigenstates of the bare system with
eigenfrequencies $\omega_{1}$ and $\omega_{2}$, respectively. The
time-dependent population amplitudes of the levels, $b_{1}(t)$ and
$b_{2} (t)$, are then governed \cite{ivanov1} by the equation
\begin{equation}
i \frac{d}{dt} b_{k}=\omega_{k}b_{k}- \Omega (t) b_{l}(t),
\label{Jxxa}
\end{equation}
where both $k$ and $l$ run the values $1, 2$ with the constrain $l
\neq k$, and $\Omega (t) = \Omega_{R} h(t)$ is the instantaneous Rabi
frequency with $\Omega_{R} = \mu \epsilon_{0}/\hbar$ being the usual
Rabi frequency as determined by the dipole transition matrix element
$\mu = \langle 1 | ez |2 \rangle $ and the electric field amplitude
$\epsilon_{0}$, while $h(t)=f(t) sin (\omega t + \phi )$ describes the
incident-field evolution with $f(t)$ having the sense of pulse shape
(for pulses of at least few cycles in duration), $\omega $ the carrier
frequency and $\phi $ the carrier-envelope offset phase. The latter is
known \cite{weg1a, weg1b, weg2, morgner, milos, paulus, gurt} to be a relevant
quantity determining the response of the system in the regime of
few-cycle pulses.

Traditionally, it has been solved in different coupling regimes either
a set of two linear differential equations for $b_{k}$ with no RWA
applied (e.g \cite{zuoa, zuob, ivanov1, ivanov2}) or more often
(e.g. \cite{mil, zuoa, zuob, levi, weg3}) the resulting set of three
linear differential equations for the Bloch vector components: $u = 2
\rm{Re}(b_{1}^{\star} b_{2})$, $v= 2 \rm{Im}(b_{1}^{\star} b_{2})$, $w=|
b_{2}|^{2} - | b_{1}|^{2}$. Instead, we prefer to work with only one
but nonlinear differential equation for the ratio $r(t) =
b_{2}(t)/b_{1}(t)$ of the population amplitudes. Through the
population conservation law, $|b_{1}|^{2}+|b_{2}|^{2}= 1$, the above
$r$ determines directly both the population inversion $w
=(|r|^{2}-1)/(|r|^{2}+1)$ and the induced dipole moment $d(t)= \langle
\psi (t) | e z |\psi (t) \rangle = \mu u =2 \mu \rm{Re}(r)/(|r|^{2}+1)$
and, consequently, the spectrum of coherently scattered light as
well. After differentiating $r$ over time and then using
Eq. (\ref{Jxxa}) one obtains \cite{genkin, parz1} $r$ to fulfil the
following differential equation:
\begin{equation}
i \frac{dr}{dt} = (r^{2}-1) \Omega (t) +\omega_{21} r,
\label{Jyxa}
\end{equation}
where $\omega_{21}$ = $\omega _{2}-\omega _{1}$ is the frequency
separation between the bare levels. This equation falls into the
family of nonlinear Riccati-type equations and a way for its iterative
solution results from the transformation
\begin{equation}
r(t)=\frac{1}{2\Omega (t)}\left(\Omega ^{eff}(t) R(t) -\omega _{21}\right),
\label{R2oa}
\end{equation}
\begin{equation}
\Omega ^{eff}(t)=\sqrt {4\Omega ^2(t)+\omega _{21}^2 },
\label{R3oa}
\end{equation}
converting Eq. (\ref{Jyxa}) into
\begin{equation}
i\frac{dR}{dt}=(R^{2}-1)\frac{\Omega ^{eff}}{2}+i\left( {\frac{\omega
_{21} }{\Omega ^{eff}} R-1} \right)\frac{\omega _{21} }{\Omega ^{eff}
\Omega } \frac{d\Omega}{dt}.
\label{R4oa}
\end{equation}
To avoid the cumbersome second term on the right-hand side, including
the derivative $d\Omega /dt$, we divide the time scale of the pulse
into a number of sufficiently narrow intervals with $t_{j}^{i} \leq
t_j \leq t_{j}^{f} $ being the running time within the $j$th
interval. In each interval of its width much shorter than half an
optical cycle we approximate the Rabi frequencies $\Omega (t_{j})$ and
$\Omega ^{eff}(t_{j})$ as constants of the values which they actually
take in the middle $(t_{j}^{m} )$ of the interval. Under such an
approximation, Eq. (\ref{R4oa}) when adapted to the $j$th interval
looks like $i (dR_{j} /dt_{j} )=(R_{j}^{2} -1) \Omega _{j}^{eff}/2$,
where $\Omega _{j}^{eff} =\Omega ^{eff}(t_{j} =t_{j}^{m} )$.  The
resulting equation has straightforward analytical solution
\begin{equation}
R_{j} (t_{j} )=\frac{1-i R_{j}^{in} \cot \left(\Omega _{j}^{eff} (t_{j}
-t_{j}^{i} )/2 \right)}{R_{j}^{in} -i \cot \left(\Omega _{j}^{eff} (t_{j}
-t_{j}^{i} )/2 \right)},
\label{R5oa}
\end{equation}
where $R_{j}^{in} =R_{j} (t_{j} =t_{j}^{i} )$ is the initial value of
$R_{j}$, i.e., that at the beginning of the $j$th interval. For the
extreme time in the interval, $t_{j} =t_{j}^{f} $, we have $R_{j}
(t_{j} =t_{j}^{f} )=R_{j+1}^{in} $, resulting in the recurrence
formula for the initial conditions
\begin{equation}
R_{j+1}^{in} =\frac{1-i R_{j}^{in} \cot \left(\Omega _{j}^{eff} (t_{j}^{f}
-t_{j}^{i} )/2 \right)}{R_{j}^{in} -i \cot \left(\Omega _{j}^{eff} (t_{j}^{f}
-t_{j}^{i} )/2 \right)}.
\label{R6oa}
\end{equation}
As a consequence of equations (\ref{R5oa}) and (\ref{R6oa}) we obtain
from Eq. (\ref{R2oa}) the solution for $r_{j}(t_{j})$
\begin{equation}
r_{j} (t_{j} )=\frac{2\Omega _{j}-\left(\omega _{21} +i\Omega
_{j}^{eff} \cot \left(\Omega _{j}^{eff} (t_{j} -t_{j}^{i} )/2 \right)
\right) r_{j}^{in} }{\omega _{21} -i\Omega _{j}^{eff} \cot
\left(\Omega _{j}^{eff} (t_{j} -t_{j}^{i} )/2\right)+ 2\Omega _{j}
r_{j}^{in} }
\label{R7oa}
\end{equation}
and also the recurrence formula for the initial conditions, $r_j^{in} $, at
the beginnings of subsequent time intervals
\begin{equation}
r_{j+1}^{in} =\frac{2\Omega_{j}-\left(\omega _{21} + i\Omega
_{j}^{eff} \cot \left(\Omega _{j}^{eff} (t_{j}^{f} -t_{j}^{i}
)/2\right) \right) r_{j}^{in} }{\omega _{21} - i\Omega _{j}^{eff} \cot
\left(\Omega _{j}^{eff} (t_{j}^{f} -t_{j}^{i} )/2 \right ) + 2\Omega
_{j} r_{j}^{in} },
\label{R8oa}
\end{equation}
where $\Omega _j^{eff} = \sqrt{4\Omega _{j}^2 +\omega _{21}^2 }$ with
$\Omega_{j}=\Omega _{R} h_{j}$ and $h_j = f(t_j^m )\sin (\omega t_j^m
+\phi )$. The solutions in the form of Eqs (\ref{R7oa}) and
(\ref{R8oa}) allow us to obtain population inversion, induced dipole
moment and electric field of coherently scattered light within the
subsequent time intervals, $t_j^i \leq t_j \leq t_j^f $, and to stick
the solutions for the intervals to cover the whole time scale of the
incident pulse.

Before writing down the final formulae it is convenient to introduce
the dimensionless strength parameter $x=\Omega _{R}/ \omega $, the
dimensionless level separation parameter $y =\omega_{21}/\omega $ and
the dimensionless time parameter $\tau = \omega t$, where $0 \leq \tau
\leq 2\pi N $ for a $N$-cycle pulse. Then, we divide each halfcycle in
the $\tau $ domain into $K$ intervals of width $\pi /K$ each, letting
$j$ to fall into the range $1\leq j \leq 2NK$.  Within the $j$th
interval, the running time covers the range $(j-1) \frac{\pi }{K}=\tau
_{j}^{i} \leq \tau _{j} \leq \tau _{j}^{f} =j \frac{\pi }{K}$ and the
middle of the interval occurs at $\tau _{j}^{m} =j \frac{\pi
}{K}-\frac{\pi }{2K}$.  Also, we make the replacement $r_{j}^{in}
=I_{j} $ and introduce the normalized effective Rabi frequency within
the $j$th interval as $x_{j}^{eff} =\Omega _j^{eff} /\omega =
\sqrt{4x_{j}^{2} +y^{2}}$, where $x_{j}= x h_{j}$ with $h_j =f(\tau
_{j}^{m}) \sin (\tau _{j}^{m} +\phi )$. In this language the recurrence
formula of Eq. (\ref{R8oa}) reads
\begin{equation}
I_{j+1} =\frac{2x_{j} -\left(y + ix_{j}^{eff} \cot \left(\pi
    x_{j}^{eff} /2K \right)\right) I_{j} }{y - ix_{j}^{eff} \cot
    \left(\pi x_{j}^{eff} /2K \right)+ 2x_{j} I_{j} }.
\label{R9oa}
\end{equation}
For a given field-system parameters ${x, y, \phi , f(\tau )}$,
Eq. (\ref{R9oa}) allows us to generate the initial conditions for all
subsequent $2NK$ time intervals from the only known initial
condition $I_{1}$ for the first interval ($I_{1}=0$ throughout this
paper). Having generated the initial conditions we calculate the
evolution of population inversion within the $j$th time interval from
\begin{eqnarray}
w_{j}(\tau_{j})=\frac{-1}{(1+|I_{j}|^{2}) (x_{j}^{eff})^{2}} \Bigl[ y
\left( y (1-|I_{j}|^{2}) + 4 x_{j} \rm{Re}(I_{j})\right) \nonumber \\
 + 4x_{j} \left( x_{j}(1-|I_{j}|^{2}) - y \rm{Re}(I_{j})\right)
\cos (x_{j}^{eff}(\tau_{j}-\tau_{j}^{i})) - 4x_{j} x_{j}^{eff}
\rm{Im}(I_{j}) \sin (x_{j}^{eff}(\tau_{j}-\tau_{j}^{i}))\Bigr],
\label{A4xa}
\end{eqnarray}
while the evolution of the induced dipole moment from
\begin{eqnarray}
d_{j}(\tau_{j})=\frac{2 \mu }{(1+|I_{j}|^{2}) (x_{j}^{eff})^{2}}
\Bigl[ x_{j} \left( y (1-|I_{j}|^{2}) +4x_{j} \rm{Re}(I_{j})
\right)  \nonumber \\ -y\left( x_{j}(1-|I_{j}|^{2}) -y
\rm{Re}(I_{j})\right) \cos(x_{j}^{eff}(\tau_{j}-\tau_{j}^{i}))
+y x_{j}^{eff} \rm{Im}(I_{j})
\sin(x_{j}^{eff}(\tau_{j}-\tau_{j}^{i}))\Bigr], \label{A5xa}
\end{eqnarray}
where $0 \leq \tau_{j}-\tau_{j}^{i} \leq \pi /K$ within each
interval. Taking the second derivative of Eq. (\ref{A5xa}) with
respect to $\tau_{j}$
results in the electric field of the coherently scattered light in the
forward direction:
\begin{eqnarray}
{\cal \epsilon} _{j}(\tau_{j}) \sim \frac{2 \mu y}{1+|I_{j}|^{2}}
\times \nonumber \\ \left[ \left(
  x_{j} (1-|I_{j}|^{2})-y \rm{Re}(I_{j}) \right)
  \cos(x_{j}^{eff}(\tau_{j}-\tau_{j}^{i})) -x_{j}^{eff}
  \rm{Im}(I_{j}) \sin(x_{j}^{eff}(\tau_{j}-\tau_{j}^{i}))\right].
\label{A6xa}
\end{eqnarray}
To study spectra we need to take Fourier transforms ($\tau _{j}
\rightarrow z$, where $z$ is the spectrometer frequency in units of
the incident light frequency $\omega $) of equations (\ref{A5xa}) and
(\ref{A6xa}) with the results
\begin{eqnarray}
d_{j}(z)=\frac{\mu e^{-iz(j-1)\pi
    /K}}{(1+|I_{j}|^{2})(x_{j}^{eff})^{2}} \Bigl[i \Bigl( x_{j} \left(
    y(1-|I_{j}|^{2}) + 4x_{j} \text{Re} (I_{j})\right) 2 f_{j}^{0}
    \nonumber\\ -y \left( x_{j}(1-|I_{j}|^{2}) -y \text{Re} (I_{j})\right)
    (f_{j}^{-1}+f_{j}^{+1})\Bigr) + yx_{j}^{eff} \text{Im} (I_{j})
    (f_{j}^{-1}-f_{j}^{+1})\Bigr]
\label{A7xa}
\end{eqnarray}
and
\begin{eqnarray}
{\cal \epsilon }_{j}(z) \sim \frac{\mu y e^{-iz(j-1)\pi
/K}}{1+|I_{j}|^{2}} \times  \nonumber \\ \left[ i \left(
x_{j}(1-|I_{j}|^{2}) - y \text{Re}(I_{j}) \right)
(f_{j}^{-1}+f_{j}^{+1}) - x_{j}^{eff} \text{Im}(I_{j})
(f_{j}^{-1}-f_{j}^{+1})\right], \label{A9xa}
\end{eqnarray}
where
\begin{equation}
f_{j}^{q}=\frac{e^{-i (z+q x_{j}^{eff}) \pi /K} -1}{z+q x_{j}^{eff}}
\label{A0xa}
\end{equation}
with $q=0, \pm 1$. Finally, to cover the whole time scale of the pulse
one needs to sum up equations (\ref{A4xa})-(\ref{A9xa}) over $j$,
taking into account equations (\ref{R9oa}) and (\ref{A0xa}).

\section{Quality of the iterative method}\label{3}

We have extensively examined the accuracy of the iterative method
    (equations (\ref{R9oa}) -- (\ref{A0xa})) in wide ranges of pulse
    shapes $f(\tau )$, pulse strengths $x$, carrier frequencies $y$
    and carrier-envelope phases $ \phi $. In any case the method was
    found to be able to fit the results of direct numerical
    integrations of the Riccati-type Eq. (\ref{Jyxa}), provided that
    $K$, i.e., the number of intervals into which we divide each
    optical halfcycle was chosen appropriately. Generally, the higher
    $K$ the better was the quality of the method, as
    expected. However, $K$ of the order of only a few units or at most
    ten appeared to be sufficient to ensure good-quality of the method
    for not too strong pulses ($x \leq 1$). For extremely strong
    pulses ($x >> 1$), generating fast population dynamics on a time
    scale of an optical cycle, an increase in $K$ was needed for the
    method to reproduce all details of the numerical
    solution. However, even in the latter case only a little of
    computer time was consumed to accomplish successfully the
    iterative procedure with the use of the short--time--interval
    analytical solutions, i.e., equations (\ref{R9oa}) -
    (\ref{A0xa}).

\begin{figure}[p] 
\centerline{\resizebox{0.6\textheight}{!}{\rotatebox{0}
{\includegraphics{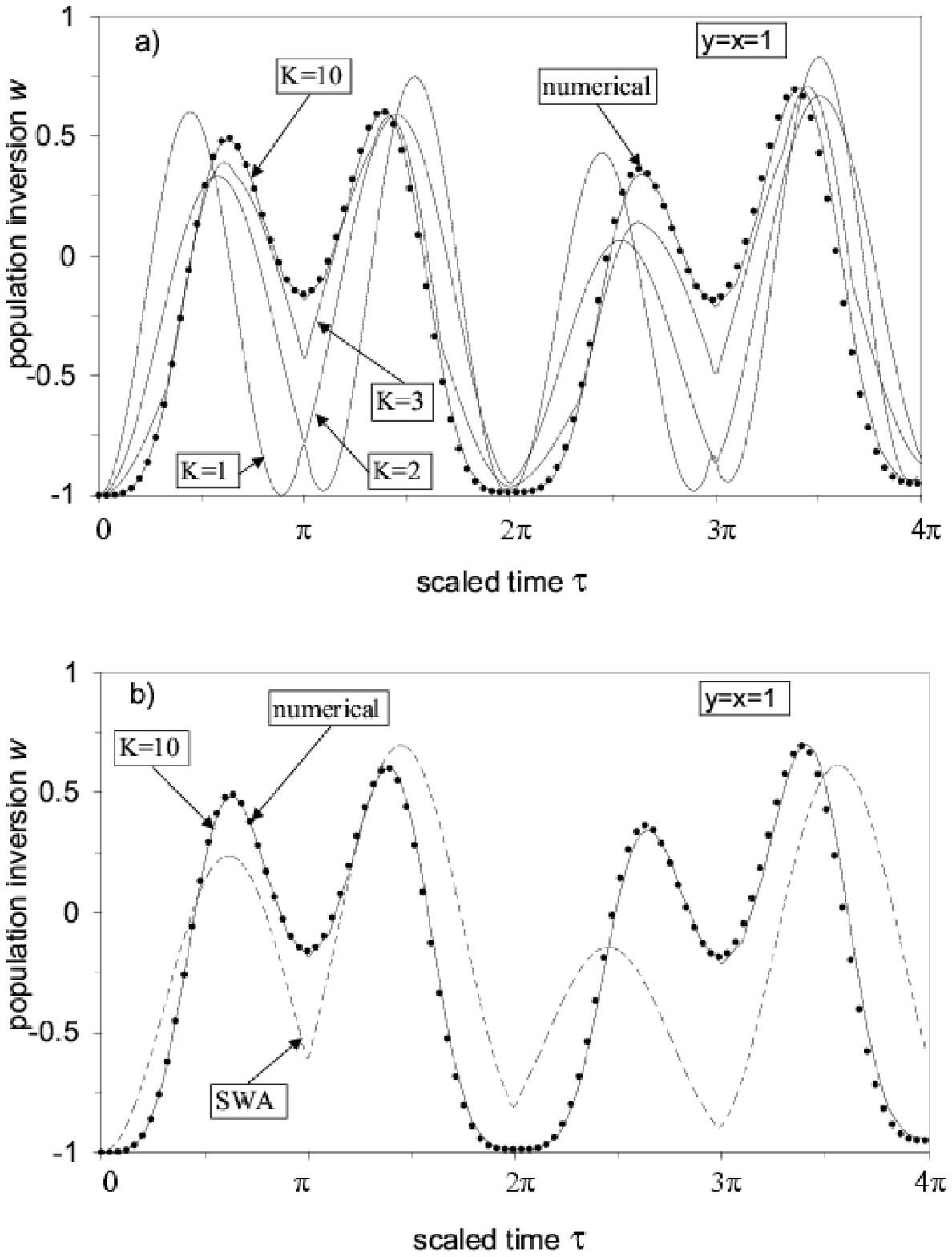}}}}
\caption{The evolution of population inversion $w$ versus
 $K$ under the box--shaped ($f(\tau )=1$) sine--like ($\phi =0$)
 two--cycle ($N=2$) pulse of $x=y=1$. Solid lines - the results of the
 iterative procedure exploiting Eq. (\ref{A4xa}), dotted lines - the
 results of direct numerical integration of the Riccati--type
 Eq. (\ref{Jyxa}) for $r$, dashed line - the results of square wave
 approximation.}
\label{fig1}
\end{figure}

To exemplify the effect of better quality of the iterative method when
    increasing $K$, let us focus on the one--photon resonance ($y=1$)
    by a pulse of moderate strength ($x=1$). We intentionally take
    this case because it is covered neither by the strong-coupling
    ($y<<1$ and $x>>y$) analytical solution of Ivanov et
    al. \cite{ivanov1, ivanov2} nor by the multiphoton-excitation
    ($y>>1$ and $x<<y$) analytical solution of Zuo et al. \cite{zuoa,
    zuob}.  Moreover, to assess the square-wave-approximation (SWA)
    solution of Tritschler et al. \cite{weg3} we choose a box-shaped
    ($f(\tau )=1$) sine-like ($\phi =0$) pulse. The SWA, originally
    applied to the system of optical Bloch equations, consisted in
    replacing the sequence of halfcycles of the electric field by the
    sequence of identical squares, each of a height ensuring the areas
    under the halfcycle and square to be equal. In terms of our
    short--time--interval analytical solution, SWA corresponds to the
    choice of $K=1$ ($h_{j}= (-1)^{j+1}$) and to rescaling $x
    \rightarrow \frac{2}{\pi} x$ resulting in $x_{j}= (-1)^{j+1}
    \frac{2}{\pi } x$. In this limit our equations for
    $w_{j}(\tau_{j})$ and $d_{j}(\tau_{j})$ convert into those of
    Tritschler et al. obtained by a different analytical approach
    exploiting the Bloch equations. For the pulse of $N=2$ cycles in
    duration, now available in the laboratory practice
    (e.g. \cite{weg3}), we show in Fig. (\ref{fig1}a) the effect of
    $K$ on the population inversion calculated iteratively with the
    use of Eq. (\ref{A4xa}) (solid lines), and compare this result
    with that obtained by integrating numerically the Riccati-type
    Eq. (\ref{Jyxa}) for $r$ (dotted line). As seen, the choice of
    $K=10$ ensures nearly perfect coincidence between the two
    approaches. On the other hand, Fig. (\ref{fig1}b) provides a
    comparison between our iterative results at $K=10$ and the SWA
    results (dashed line) leading to a conclusion that the square-wave
    approximation can be used only to general predictions of
    qualitative nature.

\begin{figure}[p] 
\centerline{\resizebox{0.6\textheight}{!}{\rotatebox{0}
{\includegraphics{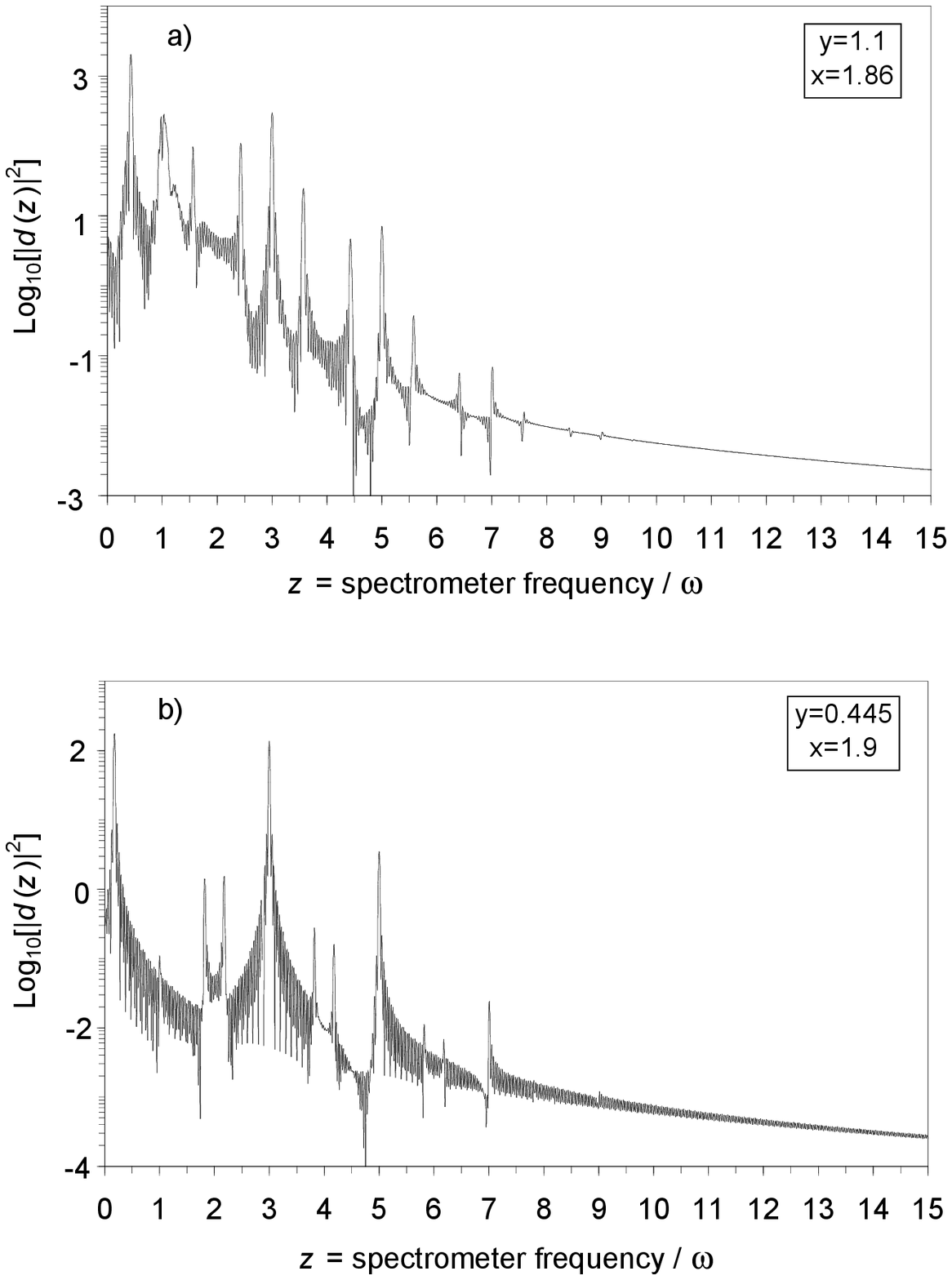}}}}
\caption{Photon--emission spectra, $|d(z)|^{2}$, from a
  model two--level molecular hydrogen ion, calculated iteratively by
  using Eq. (\ref{A7xa}). {\it (a)}: near--resonance case ($y=1.1$ and
  $x=1.86$) for the $f(\tau )\cos(\tau )$ pulse with $f(\tau )$
  gaussian increasing by $10$ optical cycles and then keeping the
  value of $1$ up to $30$ cycles; {\it (b)}: strong--coupling case
  ($y=0.445$ and $x=1.9$) for the $30$--cycle $f(\tau )\cos(\tau )$
  pulse with $f(\tau )=1$. The conditions are as those in the papers
  by Zuo et al. (Fig. 6b in \cite{zuob} and Fig. 7 in \cite{zuoa},
  respectively). For the high--resolution of the presented spectra we
  choose $K= 100$.}
\label{fig2}
\end{figure}

 To prove a good quality of our iterative method we now recover with
    it some numerically calculated spectra of light coherently
    scattered by two-level systems, available in the literature. One
    such a two-level system that has received a lot of attention in
    the past is the lowest pair of different-symmetry electronic
    levels of the $H_{2}^{+}$ molecular ion ($1\sigma_{g}$ and
    $1\sigma_{u}$), a pair being well isolated from other levels
    particularly at large internuclear distances. In particular, Zuo
    et al. show in Fig. 6b of their paper \cite{zuob} the two-level
    numerically calculated photon-emission spectrum of $H_{2}^{+}$ in
    the near-resonance region translating into our $y=1.1$ and
    $x=1.86$. The spectrum was obtained by assuming the $f(\tau
    )\cos(\tau )$ electric field with $f(\tau )$ gaussian increasing
    by $10$ optical cycles ($f(\tau ) = exp[-((\tau -20\pi )/10 \pi
    )^{2}]$ for $ 0 \leq \tau \leq 20\pi $) and then keeping a
    constant value up to $30$ cycles ($f(\tau )=1$ for $20\pi < \tau
    \leq 60\pi $). For the above conditions, we apply our
    Eq. (\ref{A7xa}) (with $\mu $ put to $1$) along with
    Eq. (\ref{R9oa}) to present in Fig. (\ref{fig2}a) the iteratively
    calculated spectrum $|d(z)|^{2}= |\sum_{j}^{} d_{j}(z)|^{2}$ with
    $1 \leq j \leq 2NK = 60K$. To achieve high resolution of our
    spectrum we chose $K=100$ and we will maintain this choice through
    all other figures to be presented. Our spectrum of
    Fig. (\ref{fig2}a) consists of Mollow triplets occurring at each
    odd-order harmonic ($1, 3, 5$ and $7$) with the same sideband
    separation within the triplets. This iteratively obtained
    structure is in full agreement with the numerical spectrum of Zuo
    et al. (Fig. 6b in \cite{zuob}). In a different paper Zuo et al.
    \cite{zuoa} give, for the $f(\tau )\cos\tau = \cos\tau $ field,
    their numerical spectrum for the same system but under the
    so-called strong-coupling conditions meaning in our notation $y =
    0.445$ and $x = 1.9$ ($x/y = \Omega_{R}/\omega_{21} =
    4.27$). Under these conditions our iterative spectrum generated
    from Eq. (\ref{A7xa}) for $N=30$ cycle pulse is shown in
    Fig. (\ref{fig2}b). An interesting feature of the iterative
    spectrum are (besides the familiar odd-order harmonics $3$, $5$
    and $7$) the doublets around the positions of even harmonics
    caused by the large Rabi splittings of the odd harmonics. This
    spectrum is a counterpart of the numerical spectrum of Zuo et
    al. (Fig. 7 in \cite{zuoa}).

\begin{figure}[p] 
\centerline{\resizebox{0.6\textheight}{!}{\rotatebox{0}
{\includegraphics{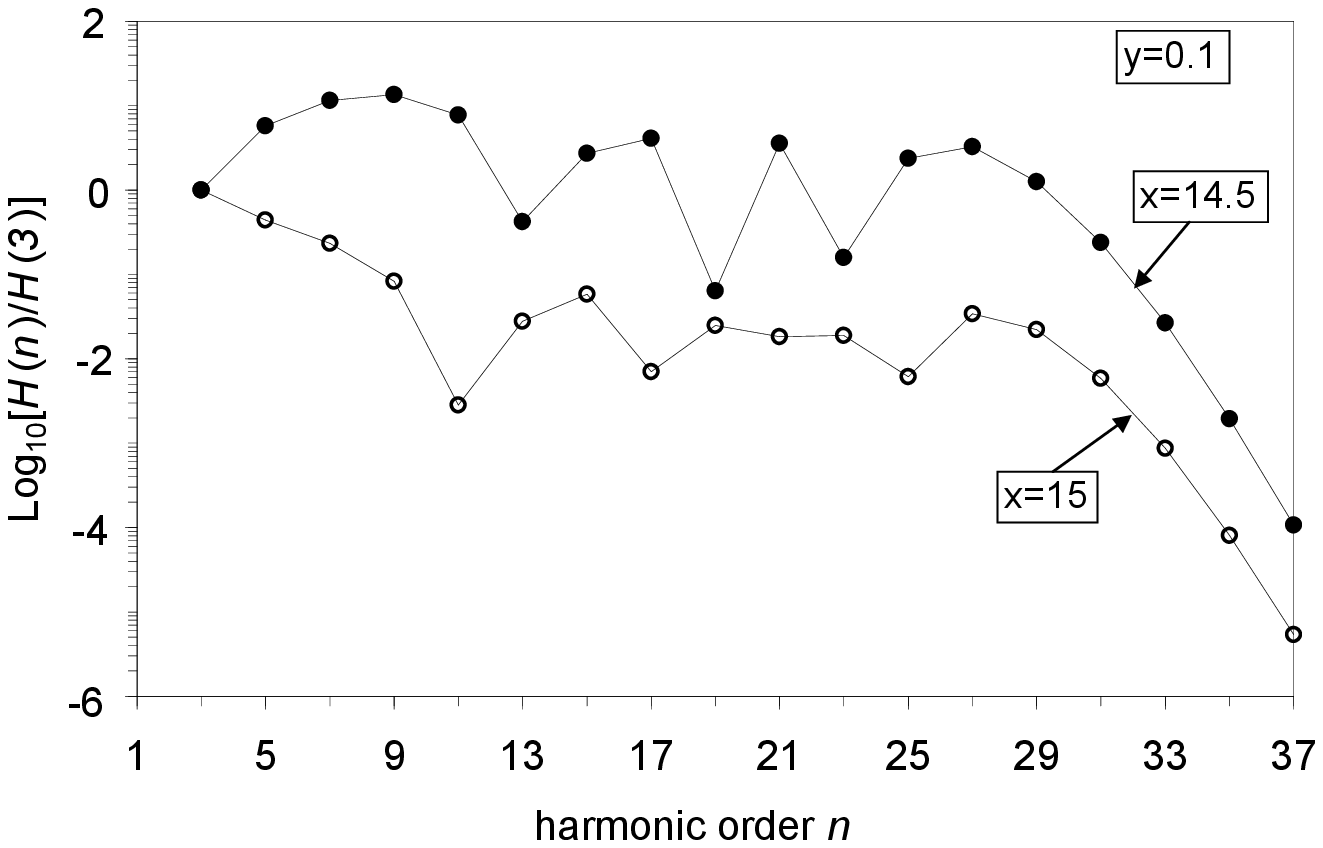}}}}
\caption{Normalized heights, $H(n)$, of the odd--harmonic
    peaks calculated iteratively with the use of Eq. (\ref{A7xa}), for
    the case of extremely strong coupling ($y<<1$, $x>>y$) of the
    two--level system to $30$--cycle pulse of $f(\tau )\cos(\tau )$
    form with $f(\tau )=1$. The conditions are as those in the paper
    by Ivanov and Corkum (Fig. 3 in \cite{ivanov1}).}
\label{fig3}
\end{figure}

Also Ivanov et al.  \cite{ivanov1, ivanov2} have calculated the
    emission spectra from molecular ions but using their analytical
    formula (Eq. (52) in \cite{ivanov1}) derived in the limiting case
    of extremely strong coupling ($y<<1$ and $x>>y$ in our
    language). We applied our Eq. (\ref{A7xa}) to this region and
    obtained with it the results shown in Fig. (\ref{fig3}). This
    figure presents the heights of the odd-harmonic peaks, $H(n)$,
    normalized to the third harmonic peak, for $y=0.1$ and two values
    of $x=14.5$ and $15$, respectively. We have assumed a $30$--cycle
    pulse of the form $f(\tau )\cos\tau $ with $f(\tau )=1$. Our
    Fig. (\ref{fig3}), obtained along the iterative procedure,
    coincides perfectly with the appropriate results of Ivanov and
    Corkum (Fig. 3 in \cite{ivanov1}).

\begin{figure}[h] 
\centerline{\resizebox{0.6\textheight}{!}{\rotatebox{0}
{\includegraphics{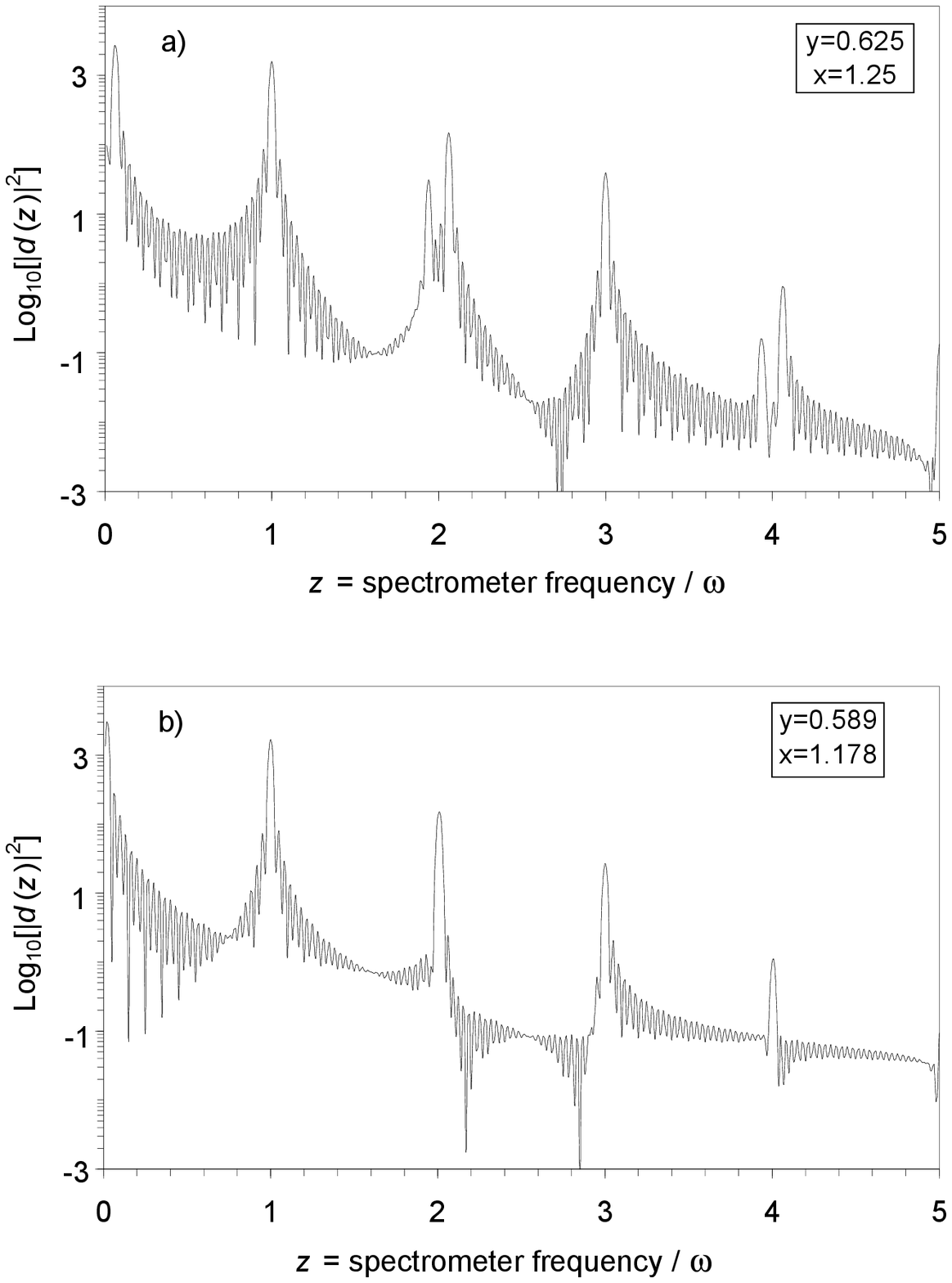}}}}
\caption{Spectra from a symmetric double--well structure
obtained iteratively by applying Eq. (\ref{A7xa}), for $30$--cycle
pulse of $f(\tau )\cos(\tau )$ type with $f(\tau )=1$ and
frequency--strength parameters $y=0.625$, $x=1.25$ ({\it (a)}) and
$y=0.589$, $x=1.178$ ({\it (b)}). The conditions are as those in the
paper by Levinson et al. (Fig. 2 in \cite{levi}).}
\label{fig4}
\end{figure}

A different place where two-level approximation has appeared to be
    reliable is a symmetric double quantum well \cite{ivanov2, bavli,
    levi, plata} extensively studied in the context of laser control
    of tunneling and symmetry breaking with strong short pulses. The
    latter effect results in the appearance of spectral peaks at the
    positions of even harmonics from the systems with inversion
    symmetry. For example, Levinson et al.(Fig. 2 in their paper
    \cite{levi}) give the spectra from the double-well structure
    obtained by integrating numerically the set of three Bloch
    equations for the $f(\tau )\cos\tau $ pulse with $f(\tau )=1$. The
    frequency-strength parameters in their numerical calculations fall
    into the strong-coupling region ($y=0.625$, $x=1.25$ in one case
    (their Fig. 2a) and $y=0.589$, $x=1.178$ in the other case (their
    Fig. 2b)). For the above two sets of frequency-strength parameters
    we show in Fig. (\ref{fig4}) our iterative spectra resulting from
    Eq. (\ref{A7xa}) for the $30$-cycle pulse assumed. The asymmetric
    doublets at the positions of the second and fourth harmonics,
    formed when using the first set of parameters
    (Fig. (\ref{fig4}a)), are seen to coalesce into single peaks when
    taking the other set (Fig. (\ref{fig4}b)). Moreover, the second
    set of parameters results in shifting the low-frequency component
    of the spectrum towards zero. Both behaviours of our iteratively
    obtained spectra are the same as those in the numerical spectra of
    Levinson et al. (Fig. 2 in \cite{levi}) and are connected with
    approaching the so-called accidental degeneracy of two Floquet
    states of the system \cite{ivanov2, bavli, plata} at some
    parameters. The parameter $x$ from the second set does nearly
    satisfy the condition of the accidental degeneracy, i.e., it
    ensures for the Bessel function $J_{0}(2x)$ to drop to zero \cite{
    ivanov2, plata}.

\begin{figure}[p] 
\centerline{\resizebox{0.6\textheight}{!}{\rotatebox{0}
{\includegraphics{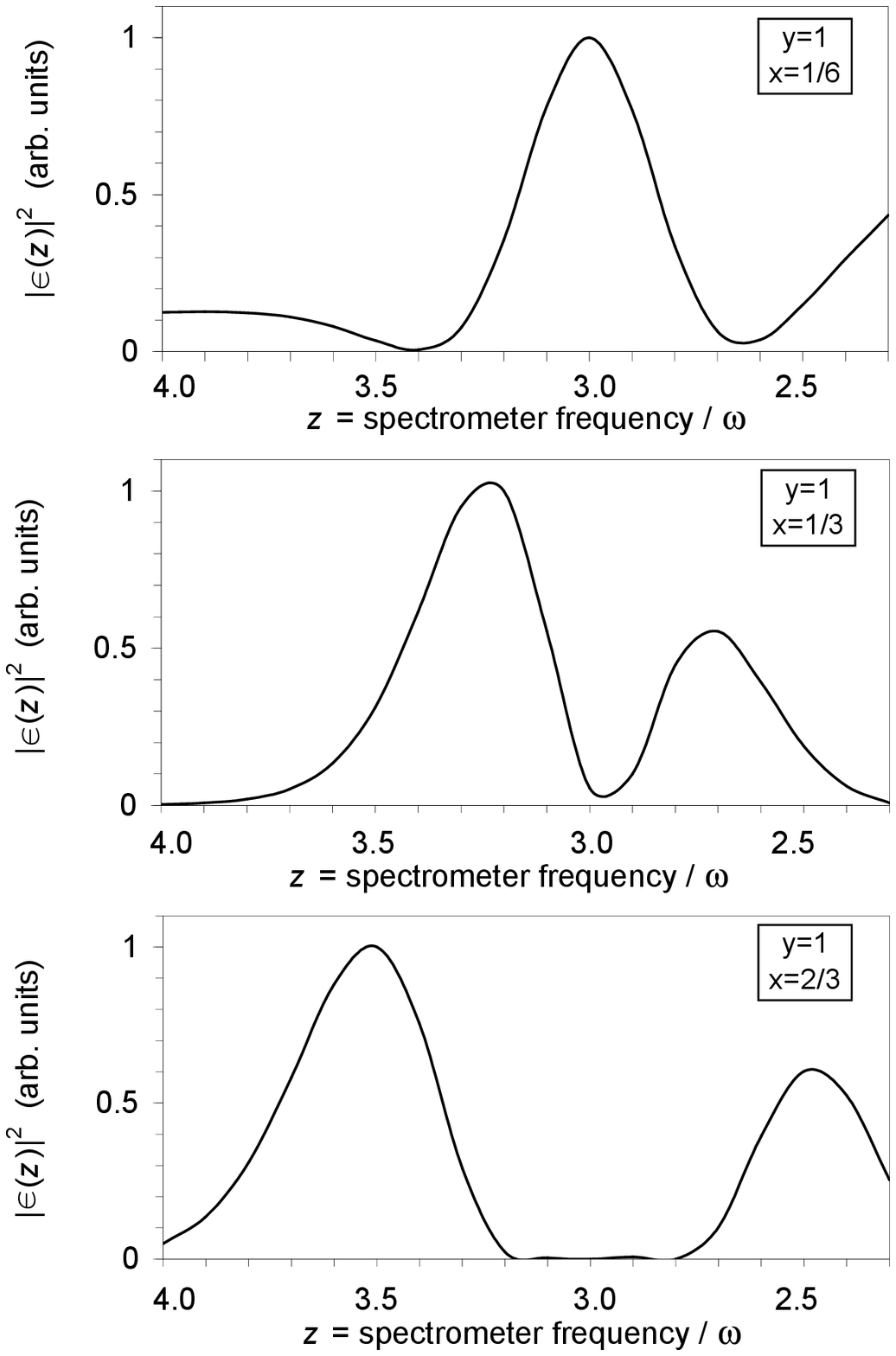}}}}
\caption{Evolution of the third--harmonic peak into a doublet with
    increasing pulse strength parameter $x$, calculated iteratively
    with the use of Eq. (\ref{A9xa}). The conditions ($5$ fs pulse of
    $\text{sech}(\tau /\tau_{0}) \cos(\tau )$ form) are close to those
    of numerical simulations by M\"ucke et al. (Fig. 2 in
    \cite{weg1a}). The positions of peaks in the doublet agree with
    those from numerical simulations but the heights are reversed.}
\label{fig5}
\end{figure}

M\"ucke et al., \cite{weg1a} have also used the two-level model to
    simulate numerically the spectra of light emitted around the third
    harmonic from $GaAs$ semiconductor exposed to $5$fs pulse of
    $\text{sech}(\tau /\tau_{0}) \cos\tau $ form, where $\tau_{0}=\tau
    _{FWHM}/1.763$. The results of their simulations (Fig. 2 in
    \cite{weg1a}) reveal the evolution of the third-harmonic peak into
    a doublet structure when increasing the envelope pulse area
    $A$. For the $\text{sech}(\tau /\tau_{0})$ envelope, the area $A$
    is related to our $x$ parameter through $A= \pi \tau_{0} x
    =(2\pi^{2}/1.763)N_{FWHM}x$, where $N_{FWHM}$ is the full width at
    half maximum (FWHM) measured in optical cycles ($N_{FWHM}=1.71$ in
    this case). For the conditions close to those of M\"ucke et al.,
    we present in Fig. (\ref{fig5}) our iterative spectra, obtained
    from Eq. (\ref{A9xa}). Our spectra are a qualitative reproduction
    of the numerical spectra of M\"ucke et al. (Fig. 2 in
    \cite{weg1a}). A possible source of only qualitative agreement in
    this case is that our spectra are the pure response of the system,
    i.e., with no propagation effects included which were naturally
    taken into account in the simulations of M\"ucke et al. by
    coupling the Bloch equations to the Maxwell equations.

\begin{figure}[p] 
\centerline{\resizebox{0.5\textheight}{!}{\rotatebox{0}
{\includegraphics{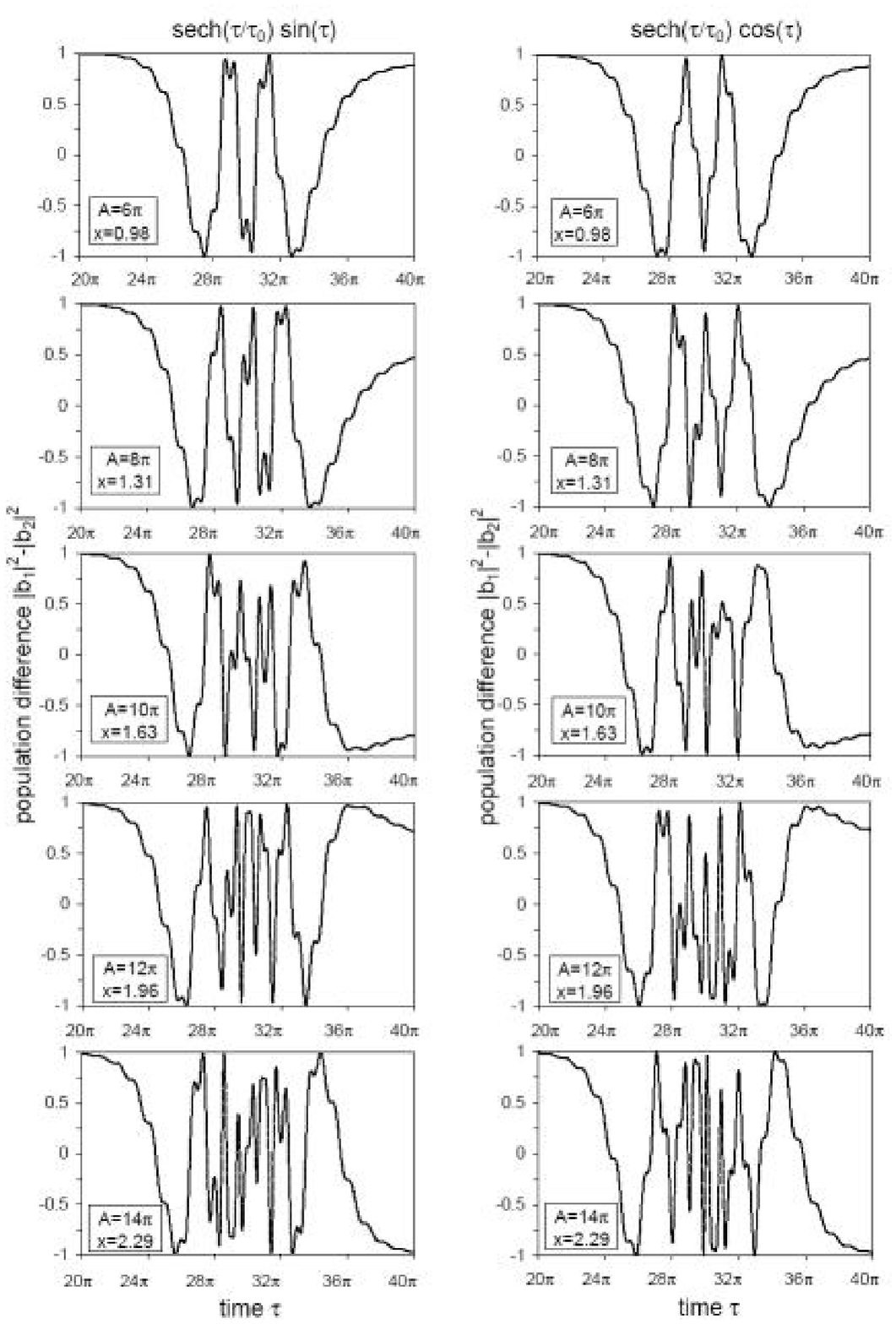}}}}
\caption{Population difference,
    $|b_{1}|^{2}-|b_{2}|^{2}$, between the ground and excited states
    versus time, for $h(\tau )=\text{sech}(\tau /\tau_{0}) \sin(\tau )$
    pulse (left--side column) and $h(\tau )=\text{sech}(\tau /\tau_{0})
    \cos(\tau )$ pulse (right--side column), {\it i.e.}, the pulses
    with their carrier--envelope phases shifted to each other by $\pi
    /2$. The graphs were obtained iteratively by using
    Eq. (\ref{A4xa}). The chosen parameters: $N_{FWHM}=1.72$, $y=1$,
    while the envelope pulse areas $A$ and the corresponding strength
    parameters $x$ are shown in the graphs. The left--column graphs
    reproduce the fully numerical results of Hughes (Fig. 3 in
    \cite{hughes1}).}
\label{fig6}
\end{figure}

\section{Carrier--envelope phase effects}\label{phase eff}

We now apply the iterative method to calculate, for a particular case,
the dependence of the two--level--system response on the phase
difference ($\phi$ ) between carrier wave and the maximum of the pulse
envelope. To be specific, we make recurrence to the carrier--wave Rabi
flopping originally studied numerically by Hughes \cite{hughes1} for a
resonant ($y=1$) pulse $h(\tau )=\text{sech}(\tau /\tau_{0}) \sin(\tau
)$ of a fixed FWHM ($N_{FWHM}=1.72$) but different pulse envelopes
$A=19.24 x$. By coupling the optical Bloch equations to the Maxwell
equations, Hughes considered propagation of the $A=2 l \pi $ pulses
through a two--level medium, where $l$ was an integer. For the areas
$A=6\pi - 14\pi $ of Hughes, the left--hand side column of
Fig. \ref{fig6} shows population differences,
$|b|_{1}^{2}-|b|_{2}^{2}$, versus time obtained by our iterative
procedure with the use of Eq. (\ref{A4xa}). Our results practically do
not differ by nothing from the original numerical results obtained by
Hughes just near the front--face of the two--level material, i.e.,
where the propagation effects were not important yet (the left--side
column of Fig. 3 in \cite{hughes1}). Our graphs confirm the original
result of Hughes on incomplete Rabi flops at $A\geq 8\pi $. On the
other hand, the right--side column of our Fig. \ref{fig6} shows our
iteratively obtained population differences but for the $h(\tau )=
\text{sech}(\tau /\tau_{0})\cos(\tau )$ pulse, i.e., the pulse with its
carrier--envelope phase $\phi $ shifted by $\pi /2$ with respect to
the pulse used by Hughes. Some differences caused by this shift are
clearly seen in the middle parts of the population difference
curves. These parts correspond to the times for which the pulse intensity
has already evolved to its high values. The main differences
introduced by changing the carrier--envelope phase $\phi $ consist in
either converting the double peaks into single ones (and vice versa)
or inverting the asymmetry in double peaks.

\begin{figure}[p] 
\centerline{\resizebox{0.6\textheight}{!}{\rotatebox{0}
{\includegraphics{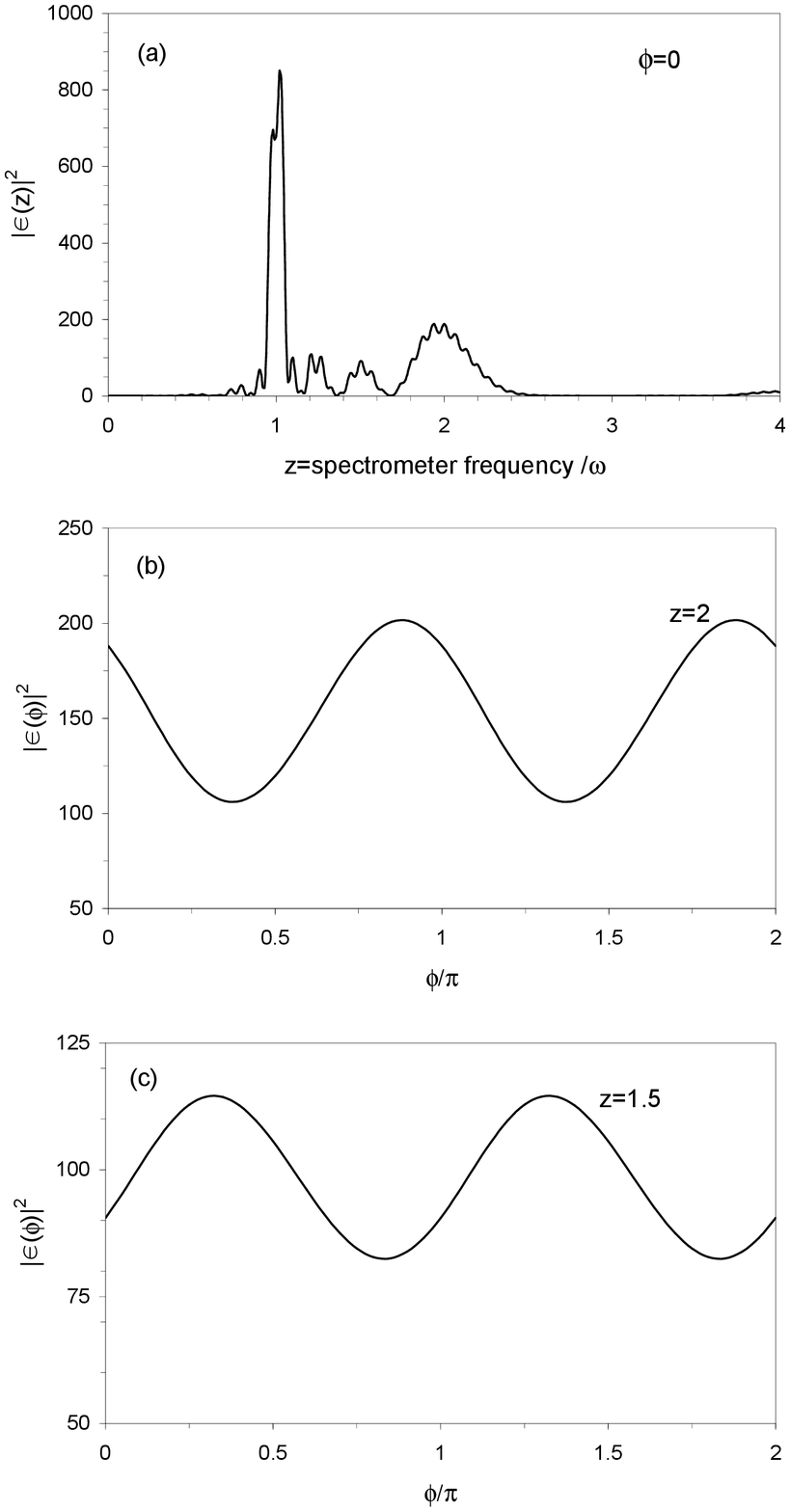}}}}

\caption{ $(a)$ The iteratively calculated (from
  Eq. (\ref{A9xa})) photon emission spectrum generated by $h(\tau
  )=\text{sech} (\tau /\tau_{0}) \sin(\tau + \phi ) $ pulse of the
  parameters $\phi =0$, $N_{FWHM}=1.72$, $y=1$ and $x=1.31$ ($8\pi $
  pulse). $(b)$ The calculated (from Eq. (\ref{A9xa})) height of
  the spectral peak at the position of second harmonic ($z=2$) versus
  carrier--envelope offset phase $\phi $. The $\phi $--dependence with
  a period of $\pi $ agrees with the results of fully numerical
  calculations by other Authors \cite{weg1b, weg2} for different pulse
  shapes and light--matter parameters. $(c)$ The same as (b) but
  for the small peak at $z=1.5$ in (a).}
\label{fig7}
\end{figure}

The above $\phi $--sensitivity of population inversion produces the
dependence of the spectrum of scattered light on carrier--envelope
phase. In Fig. 7a, we show the spectrum calculated
iteratively with the use of Eq. (\ref{A9xa}) for the Hughes pulse
$h(\tau )=\text{sech}(\tau /\tau_{0})\sin(\tau + \phi )$, i.e., with
$\phi =0$, $N_{FWHM}=1.72$, $y=1$ and $x=1.31$ (this $x$ corresponds
to the envelope pulse area $A=8\pi $). Except the spectral peak at the
fundamental frequency ($z=1$), one sees a well pronounced peak at the
position of second harmonic ($z=2$) because the chosen $x$ is in the
vicinity of the value (1.178) ensuring the accidental degeneracy
($J_{0}(2x)=0$) of the Floquet states of the system (compare
Fig. \ref{fig4} and its discussion). A similar peak around $z=2$ was
found by Tritschler et al. (Fig. 1a in \cite{weg2}) on the basis of
their numerical solution of the two--level Bloch equations for a
different pulse envelope ($\text{sinc}(\tau /\tau_{0})$ instead of
$\text{sech}(\tau /\tau_{0})$) and different light--matter parameters
($N_{FWHM}=1.81$, $y=2$, $x=0.76$). In addition to Fig. 7a,
we show in Fig. 7b the dependence (calculated iteratively
from Eq. (\ref{A9xa})) of the height of the spectral peak at $z=2$ on
the carrier--envelope phase $0\leq \phi \leq 2\pi $ in the pulse
$h(\tau )=\text{sech}(\tau /\tau_{0})\sin(\tau + \phi )$. The $\phi
$--dependence is well seen and has a period of $\pi $ in agreement
with fully numerical calculations of Tritschler et al. \cite{weg2} and
M\"ucke et al. (Fig. 1b in \cite{weg1b}) exploiting the optical Bloch
equations. The same periodicity is seen in Fig. 7c
corresponding to the small peak in Fig. 7a around $z=1.5$.

\section{Summary}\label{concl}

On the basis of a nonlinear Riccati-type equation, analytically solved
in very short time intervals (shorter than half the optical period), we
have formulated an effective iterative procedure for the problem of a
two-level system exposed to a linearly polarized electromagnetic
pulse. For different light-matter couplings (from weak through
moderate to strong ones), we have proved very good quality of the
procedure by recovering with it a number of subtle effects met in the
previous numerically calculated photon-emission spectra and population
inversion.  We have applied the procedure developed to describe some
carrier-envelope phase effects in extreme nonlinear optics,
particularly in population inversion and spectrum of coherently
scattered light. In the regime of a few--cycle pulses, these
carrier--envelope phase effects are of current interest \cite{morgner,
milos, paulus, gurt, sans, bra}. 

If necessary, one could reinterpret the interaction Hamiltonian
appropriately thus making the iterative procedure applicable to any
form of time--dependent coupling within a two--level system.

\end{document}